\documentclass[aps,pra,twocolumn,superscriptaddress,nofootinbib,floatfix,longbibliography,10pt]{revtex4-2}

\usepackage{amsmath}
\usepackage{amssymb}
\usepackage{amsthm}
\usepackage{dsfont}
\usepackage[bookmarksnumbered=true,hypertexnames=false,colorlinks=true,linkcolor=blue,urlcolor=blue,citecolor=blue,anchorcolor=blue]{hyperref}
\usepackage[ruled,vlined]{algorithm2e}
\usepackage{algorithmic}
\usepackage{physics}
\usepackage{tikz}
\usepackage{graphicx}
\usepackage[caption=false,justification=raggedright,font={sf},singlelinecheck=false,position=top]{subfig}

\newtheorem{theorem}{Theorem}

\begin{document}

\title{Counting with the quantum alternating operator ansatz}

\author{Julien Drapeau}
\email[Corresponding author: ]{julien.drapeau@usherbrooke.ca}
\affiliation{Départment de physique, Université de Sherbrooke, Sherbrooke, Québec, J1K 2R1, Canada}
\affiliation{Institut quantique, Université de Sherbrooke, Sherbrooke, Québec, J1K 2R1, Canada}

\author{Shreya Banerjee}
\affiliation{Départment de physique, Université de Sherbrooke, Sherbrooke, Québec, J1K 2R1, Canada}
\affiliation{Institut quantique, Université de Sherbrooke, Sherbrooke, Québec, J1K 2R1, Canada}

\author{Stefanos Kourtis}
\affiliation{Départment de physique, Université de Sherbrooke, Sherbrooke, Québec, J1K 2R1, Canada}
\affiliation{Institut quantique, Université de Sherbrooke, Sherbrooke, Québec, J1K 2R1, Canada}
\affiliation{Départment d'informatique, Université de Sherbrooke, Sherbrooke, Québec, J1K 2R1, Canada}

\begin{abstract}
	We introduce a variational algorithm based on the quantum alternating operator ansatz (QAOA) for the approximate solution of computationally hard counting problems. Our algorithm, dubbed VQCount, is based on the equivalence between random sampling and approximate counting and employs QAOA as a solution sampler. We first prove that VQCount improves upon previous work by reducing exponentially the number of samples needed to obtain an approximation within an arbitrary small multiplicative factor of the exact count. Using tensor network simulations, we then study the typical performance of VQCount with shallow circuits on synthetic instances of two \#P-hard problems, positive \#NAE3SAT and positive \#1-in-3SAT. We employ the original quantum approximate optimization algorithm version of QAOA, as well as the Grover-mixer variant which guarantees a uniform solution probability distribution. We observe a tradeoff between QAOA success probability and sampling uniformity, which we exploit to achieve an empirical efficiency gain over both naive rejection sampling and Grover-based quantum counting. Our results highlight the potential and limitations of variational algorithms for approximate counting.
\end{abstract}

\maketitle

\section{Introduction}

By leveraging classical and quantum resources, variational quantum algorithms (VQAs)~\cite{cerezoVariationalQuantumAlgorithms2021} perform specific computational tasks without requiring error-corrected qubits. Having drawn considerable attention in recent years, these algorithms are now routinely applied to prevalent combinatorial optimization problems, such as finding maximum cuts or independent sets in graphs, attempting to compete against classical algorithms refined over decades. This effort has yielded considerable improvements in both VQAs and classical algorithms for a variety of challenging problems. Yet, without definitive proofs or scalable demonstrations of quantum speedups, a conclusive answer as to whether VQAs can truly offer a quantum advantage is still pending.

Recent works introduce strategies that refine VQAs to better exploit the structure of the problems studied. Examples include appropriately restricting the variational parameter space~\cite{sackQuantumAnnealingInitialization2021} or the qubit Hilbert space~\cite{hadfieldQuantumApproximateOptimization2017}. These approaches aim to make it easier for VQAs to solve a given problem, thereby seeking to outperform the state-of-the-art classical algorithms for the same problem. Here we adopt a complementary approach, which is to turn to problems that are intrinsically harder than combinatorial optimization for classical algorithms but perhaps not for VQAs. Our strategy is similar in spirit to boson sampling~\cite{aaronsonComputationalComplexityLinear2011} and random circuit sampling~\cite{boulandComplexityVerificationQuantum2019}, tasks designed to be hard for classical computers. However, both these tasks are unlikely to have an impact on real-world applications.

These considerations motivate us to turn our attention to counting problems. This class of problems asks \emph{how many} solutions there are for a given instance of a decision problem. Counting problems lie higher in the computational complexity hierarchy than the optimization problems commonly targeted with near-term quantum algorithms. Counting problems arise across many disciplines, with applications in probabilistic reasoning~\cite{rothHardnessApproximateReasoning1996, sangPerformingBayesianInference2005, abramsonHailfinderBayesianSystem1996}, network reliability~\cite{valiantComplexityEnumerationReliability1979, duenas-osorioCountingBasedReliabilityEstimation2017}, statistical physics~\cite{jerrumPolynomialtimeApproximationAlgorithms1993}, and artificial intelligence~\cite{balutaQuantitativeVerificationNeural2019}. For decision problems whose solutions are verifiable in time that scales polynomially with the size of the problem instance, that is, in polynomial time, the corresponding task of counting solutions belongs to complexity class \textsf{\#P}. The class \textsf{\#P}, like \textsf{NP}, contains a subset of hard problems which are unlikely to admit exact polynomial-time solutions. Naturally, the counting counterpart of every \textsf{NP}-hard problem in \textsf{NP} is \textsf{\#P}-hard. In the quantum domain, counting was originally formulated as an application of amplitude estimation~\cite{brassardQuantumAmplitudeAmplification2002, wieSimplerQuantumCounting2019, aaronsonQuantumApproximateCounting2020}, which uses the Grover iteration~\cite{groverFastQuantumMechanical1996} as a subroutine. More recently, VQAs based on the quantum alternating operator ansatz (QAOA)~\cite{hadfieldQuantumApproximateOptimization2019} for (weighted) counting~\cite{sundarQuantumAlgorithmCount2019} and enumeration~\cite{zhangGroverQAOA3SATQuadratic2024} were introduced. These algorithms are parameterized by the success rate of QAOA circuits, i.e., the probability of a measurement yielding a solution.

The improbability of exact polynomial-time algorithms for \textsf{\#P}-hard problems has spurred extensive research into practical solvers~\cite{biereHandbookSatisfiability2021}. Exact counting heuristics employ techniques such as component caching~\cite{sangCombiningComponentCaching2004}, knowledge compilation~\cite{lagniezImprovedDecisionDNNFCompiler2017}, and tensor networks~\cite{kourtisFastCountingTensor2019, dudekEfficientContractionLarge2020} to outperform brute-force enumeration, but they typically remain limited to instances with at most a few hundred variables. Approximate counting algorithms, which sacrifice exactness for performance, fall into two main categories. Hashing-based methods recursively partition the solution space with random hash functions and query an NP oracle, such as a SAT solver, to test for emptiness~\cite{stockmeyerComplexityApproximateCounting1983, chakrabortyScalableApproximateModel2013}. Sampling-based methods, by contrast, rely on generating solutions (almost) uniformly at random. An example is the celebrated algorithm by Jerrum, Valiant, and Vazirani (JVV)~\cite{jerrumRandomGenerationCombinatorial1986}, which establishes an equivalence between approximate counting and random sampling of solutions. The JVV algorithm approximately solves \textsf{\#P}-hard problems that are self-reducible \emph{if} solutions can be sampled sufficiently uniformly (the term ``self-reducible'' will be defined in Sec.~\ref{sec:jvv-algorithm}).

In this work, we introduce VQCount, an algorithm for approximate counting with VQAs. Concretely, we show how QAOA can be used as a subroutine within the JVV algorithm to build an approximate solver for self-reducible \textsf{\#P} problems. If QAOA succeeds in sampling solutions sufficiently close to uniformly with success rate $r$, VQCount requires $O\left(\frac{n^2 \log(1/\delta)}{r \varepsilon^2}\right)$ samples to obtain an approximate count to any desired multiplicative accuracy $\varepsilon$ with confidence $1-\delta$ for a problem instance of $n$ variables. Our approach then offers an exponential improvement over previous related work on approximate counting with VQAs~\cite{sundarQuantumAlgorithmCount2019}.  The key question then becomes: under which conditions can QAOA function as an efficient random sampler of solutions?

To investigate this, we evaluate the success rate and total variation distance from the uniform distribution of the output state as a function of QAOA circuit depth for a \textsf{\#P}-hard problem (\#NAE3SAT) in two limiting cases. On the one hand, the quantum approximate optimization algorithm~\cite{farhiQuantumApproximateOptimization2014} requires shallower circuits to reach solutions, but samples solutions non-uniformly. On the other hand, the Grover-mixer variant of QAOA (GM-QAOA)~\cite{bartschiGroverMixersQAOA2020} guarantees perfect uniformity but requires deeper circuits to reach solutions reliably. We therefore establish the existence of a tradeoff between sampling efficiency and sampling uniformity with respect to counting accuracy. Finally, we restrict ourselves to shallow QAOA circuits suitable for near-term applications on noisy hardware and estimate the number of shots required to obtain an approximate count within a multiplicative factor of the exact number of solutions for the \textsf{\#P}-hard problem positive \#1-in-3SAT. We observe an exponential reduction in the number of samples compared to naive rejection sampling of solutions, though the overall scaling is worse than state-of-the-art exact classical algorithms for this problem~\cite{kourtisFastCountingTensor2019,dudekEfficientContractionLarge2020,dudekParallelWeightedModel2021}.

This paper is organized as follows. Sec.~\ref{sec:qaoa} introduces the quantum alternating operator ansatz, with an emphasis on the quantum approximate optimization algorithm and its Grover-mixer variant, in the context of approximate counting. The JVV algorithm is summarized in Sec.~\ref{sec:jvv-algorithm}. With this theory, Sec.~\ref{sec:vqcount} defines the VQCount algorithm. Sec.~\ref{sec:solving-sharp-P-complete-problems-with-vqcount} presents numerical evidence of the performance of VQCount. We first benchmark the performance of VQCount on $\#$NAE3SAT and then push simulations to a higher number of qubits for the $\#$1-in-3SAT problem.

\section{Quantum alternating operator ansatz}
\label{sec:qaoa}

Variational quantum algorithms (VQAs) are formulated in terms of a parameterized quantum circuit (PQC) and a classical optimization procedure. A combinatorial optimization problem, often from the \textsf{NP}-hard complexity class, is mapped to a cost function so that solutions to the problem minimize the cost function. In a VQA, an initial easy-to-prepare quantum state is unitarily evolved with a PQC and the expectation value of the cost function is estimated by multiple measurements of the output of the circuit in an appropriate basis. One then optimizes the parameters of the PQC to minimize the cost and therefore output a state which is close to a superposition of solutions to the problem.

The quantum approximate optimization algorithm (QAOA)~\cite{farhiQuantumApproximateOptimization2014} was recognized as a promising VQA. QAOA is defined through a problem Hamiltonian $H_P$, which is diagonal in the computational basis of $n$ qubits, and a ``mixer'' Hamiltonian $H_D$ that contains off-diagonal elements. These Hamiltonians respectively define the problem operator $U_P (\gamma) = e^{-i \gamma H_P}$ and the mixer operator $U_D (\beta) = e^{-i \beta H_D}$. $U_P$ represents a phase rotation parameterized by $\gamma$ of computational basis states according to their energy with respect to $H_P$. $U_D$, parameterized by $\beta$, superimposes computational basis states which have previously acquired different phase factors, thereby leading to interference. A common choice for the drive is
\begin{equation}
	H_D^X = \sum_{i=1}^n X_i \,, \label{eq:x-mixer}
\end{equation}
where $X_i$ is the Pauli $X$ operator on the $i$-th qubit. Problem Hamiltonians are typically sums of products of Pauli $Z$ operators, such as Ising models.

The product $U_D U_P$ is applied $p$ times to an initial state $\ket{\psi_0}$, prepared to be an eigenstate of $H_D$. For example, $\ket{\psi_0} = \ket{+}^{\otimes n}$ for the drive~\eqref{eq:x-mixer}. This yields
\begin{equation}
	\label{eq:final-state}
	|\psi(\vec{\beta}, \vec{\gamma})\rangle=\underbrace{U_D(\beta_p) U_P(\gamma_p) \cdots U_D(\beta_1) U_P(\gamma_1)}_{p \text { times }} \left|\psi_{0}\right\rangle ,
\end{equation}
where $\vec{\beta} = (\beta_1, \dots, \beta_p)$ and $\vec{\gamma} = (\gamma_1, \dots, \gamma_p)$.

In the $p\to\infty$ limit, the convergence of $|\psi(\vec{\beta}, \vec{\gamma})\rangle$ to the exact solution of the problem encoded by $H_P$ is guaranteed, since $(\vec{\beta},\vec{\gamma})$ can be predetermined to model trotterized adiabatic quantum optimization~\cite{farhiQuantumApproximateOptimization2014}. Clearly, this limit is not interesting for practical use.

On the other hand, QAOA circuits for finite (preferably small) $p$ are readily implementable, at least in principle, on near-term quantum processors and have therefore been intensively investigated recently. In this case, the optimal values of $(\vec{\beta},\vec{\gamma})$ are not a priori known and a classical optimizer is employed to tune the parameters to minimize (or maximize, depending on the problem) the energy expectation
\begin{equation}
	E_{P}(\vec{\beta}, \vec{\gamma})=\left\langle\psi(\vec{\beta}, \vec{\gamma})|H_P| \psi( \vec{\beta}, \vec{\gamma})\right\rangle \,. \label{eq:Eexp}
\end{equation}
Since $H_P$ is typically a sum of Pauli string operators, this expectation can be evaluated efficiently via measurement of the circuit output~\cite{nielsenQuantumComputationQuantum2011}.

Initial interest in QAOA inspired its generalization to the quantum alternating operator ansatz, also abbreviated QAOA, in which an initial state is evolved with two singly parameterized circuits in alternation, though without necessarily implementing Hamiltonian evolution~\cite{hadfieldQuantumApproximateOptimization2019}. An instance of this generalization is the so-called Grover-mixer QAOA (GM-QAOA)~\cite{bartschiGroverMixersQAOA2020}. GM-QAOA incorporates elements of the Grover algorithm. In GM-QAOA, $U_P$ remains the same as in the original QAOA, while the drive becomes
\begin{equation}
	U_D^{\mathrm{Grover}} = U_{S}\left[ \mathds{1} -\left(1-e^{-i \beta}\right) (|0\rangle\langle 0|)^{\otimes n} \right] U_{S}^{\dagger} \,,
\end{equation}
where $U_S$ is a state preparation unitary and $\ket{\psi_0} = U_{S} \ket{0}^{\otimes n}$. The purpose of $U_S$ is to restrict QAOA within a ``feasible'' subspace of the Hilbert space by eliminating subspaces that are known a priori not to contain solutions.

The key feature here is that the amplitudes of all solutions in $|\psi(\vec{\beta}, \vec{\gamma})\rangle$ prepared by GM-QAOA are guaranteed to be \emph{exactly equal}. The importance of this feature for adapting QAOA to counting problems will become apparent below.

After initialization, a classical continuous optimization method, such as stochastic gradient descent, is used to search for (near-)optimal parameter values that minimize the expectation~\eqref{eq:Eexp}. Upon convergence, the state at the output of the quantum circuit is approximately a superposition of solutions to the problem encoded in $H_P$. A measurement in the computational basis then samples one solution from the superposition.

Given the promise of QAOA in solving optimization problems, it is reasonable to ask whether it can be adapted to counting problems. However, the application of QAOA to counting problems with \emph{exponentially many} solutions, which are the hard cases for this class, is not straightforward, for three reasons. First, QAOA circuits must produce a large separation in magnitudes between solution and non-solution amplitudes in the output state for the algorithm to succeed in producing even one solution after a tractable number of repetitions. Second, even if non-solution amplitudes vanish, if the amplitudes of a sizable subset of the solutions are significantly suppressed compared to more prominent amplitudes, then measurement of the suppressed solutions incurs a large (potentially exponential) overhead in repetitions. Finally, even if all solutions have roughly equal amplitudes and non-solutions are not present in the output state, an expected exponential number of measurements is required to naively enumerate all of them exhaustively.

These complications have hampered previous work in solving counting problems by adapting quantum optimization techniques~\cite{sundarQuantumAlgorithmCount2019}, resulting in methods requiring superpolynomial numbers of measurements even in the ideal third scenario described above.

\section{Jerrum-Valiant-Vazirani algorithm}
\label{sec:jvv-algorithm}

To overcome the aforementioned limitations of VQAs for counting, we employ a well-known result in theoretical computer science, namely, the correspondence between approximate counting and random sampling of combinatorial structures~\cite{sinclairApproximateCountingUniform1989}. This correspondence was firmly established in the work of Jerrum, Valiant, and Vazirani (JVV)~\cite{jerrumRandomGenerationCombinatorial1986} (see also~\cite{broderHowHardIt1986}), who proved the existence of an approximate algorithm that counts the solutions to a self-reducible problem \emph{iff} solutions can be sampled sufficiently uniformly. Informally, a problem is self-reducible if its solutions can be built out of solution sets of smaller instances of the same problem.

Let us consider the Boolean satisfiability (or SAT) problem as an example. A SAT instance amounts to asking whether a given Boolean formula $\varphi(x)$, where $x = x_n x_{n-1} \dots x_1$ is a bit string and $x_i = 0,1,\, i=1,\dots,n$ are Boolean variables (we identify \textsc{\texttt{True}} $\leftrightarrow$ 1, \textsc{\texttt{False}} $\leftrightarrow$ 0 for convenience), evaluates to 1. Formulas are commonly expressed in conjunctive normal form (CNF), that is, a conjunction (logical \textsc{\texttt{AND}}) of clauses, each composed of a disjunction (logical \textsc{\texttt{OR}}) of literals. A literal is defined as either a variable or its negation. SAT is a prototypical \textsf{NP}-complete problem.

\begin{figure}[t!]
	\centering
	\includegraphics[width=0.9\columnwidth]{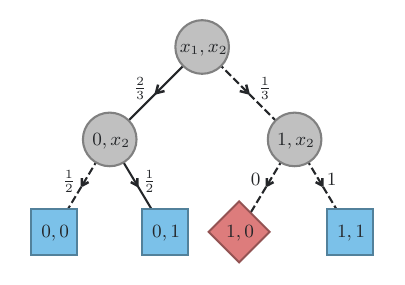}
	\caption{An example of the self-reducibility tree in the JVV algorithm for a CNF formula $\varphi(x_1, x_2)$. Vertices represent literal assignments in $\varphi$, with solutions depicted as blue squares and non-solutions as red rhombuses. Directed edges indicate the assignment of a literal along with its associated probability. Following the path of maximal probabilities, shown with a solid line, the count is obtained as $N = \frac{1}{P_{0}} \cdot \frac{1}{P_{01}}$ = 3.}
	\label{fig:jvv-algorithm}
\end{figure}

Self-reducibility is itself a research area and its general definition is intricate~\cite{hemaspaandraPowerSelfReducibilitySelectivity2020}. In the case of SAT problems that interest us here, the relevant notion of self-reducibility is expressed simply as the following equivalence:
\begin{align}
	 & \varphi(x)=1 \iff                                                                        \\
	 & (\varphi(x_1 = 0, x_{\bar{1}}) = 1) \vee (\varphi(x_1 = 1,x_{\bar{1}}) =1) \,, \nonumber
\end{align}
where $x_{\bar{1}}=x_n x_{n-1} \dots x_2$. Since variables can be arbitrarily relabeled, the choice of $x_1$ is immaterial. Also, since the new formulas on the right-hand side of the equivalence are of the same type as the one on the left-hand side, they can be similarly decomposed, leading to a recursion. The self-reducibility of SAT can thus be visualized as a rooted directed binary tree, like the one shown in Fig.~\ref{fig:jvv-algorithm}, in which each directed edge represents an assignment of a literal and each vertex represents the sub-formula that corresponds to the preceding assignment.

The total variation distance (TVD) between two discrete probability distributions $P$ and $Q$ over a set $\cal X$ is
\begin{equation}
	\label{eq:tvd}
	||P-Q||_{TV} \equiv \frac{1}{2} \sum_{x\in \mathcal{X}} |P(x) - Q(x)| \,.
\end{equation}
We define the nonuniformity $\eta$ of a distribution $Q$ as its TVD from the uniform distribution $U$:
\begin{equation}\label{eq:nonuniformity}
	\eta(Q) \equiv ||U-Q||_{TV} \equiv \frac{1}{2} \sum_{x\in \mathcal{X}} |U(x) - Q(x)| \,.
\end{equation}
We say that a generator of solutions for a given problem has nonuniformity $\eta$ if it produces samples from a distribution $Q$ over the solution set with nonuniformity $\eta(Q)$.

Let us now illustrate the link between self-reducibility, sampling, and counting established by JVV. Suppose the exact solution count for an SAT formula on $n$ variables is $N$. Denote $x_{:m}$ the prefix of length $m$ of bit string $x$ and $N_{x_{:m}}$ the number of solutions with prefix $x_{:m}$. Further, suppose that we have a generator that samples the solution set uniformly at random, such that each solution is output with probability $p^\star=1/N$ (i.e., we assume perfect uniformity for the moment). For any given solution bit string $z$, we can write
\begin{align}
	p^\star = \frac1N = & {\ } \frac{N_{z_{:1}}}{N} \cdot \frac{N_{z_{:2}}}{N_{z_{:1}}} \cdots \frac{N_{z_{:n}}}{N_{z_{:n-1}}} \\
	=                   & {\ } p(z_{:1}) \cdot p(z_{:2}|z_{:1}) \cdots p(z_{:n}|z_{:n-1})                                      \\
	=                   & {\ } p(z_{:1}) \prod_{i=1}^{n-1} p(z_{:i+1}|z_{:i}) \,,
\end{align}
where $N_{z_{:n}}\equiv1$. The conditional probability $p(z_{:i+1}|z_{:i})\equiv N_{z_{:i+1}} / N_{z_{:i}}$ is the probability that a sampled solution $z'$ has $z'_{i+1}=z_{i+1}$ given that $z'_{:i} = z_{:i}$. The above setup is sketched in Fig.~\ref{fig:jvv-algorithm} for a short CNF formula.

The JVV algorithm for approximate counting yields a multiplicative approximation to $p^\star$, and hence $N$, by approximating the conditional probabilities $p(z_{:i+1}|z_{:i})$. This is done by generating a sample set $\cal S$ and approximating the conditional probabilities by
\begin{equation}
	\tilde p(z_{:i+1}|z_{:i}) = \frac{ \tilde N_{z_{:i+1}} }{ \tilde N_{z_{:i}} } \,,
\end{equation}
where $\tilde N_{z_{:i}}$ is the number of solutions \emph{in the sample set} that begin with $z_{:i}$. That is,
\begin{equation}
	\tilde p^\star = \tilde p(z_{:1}) \prod_{i=1}^{n-1} \tilde p(z_{:i+1}|z_{:i}) \approx \frac1N \,.
\end{equation}
The key element in the proof of the JVV algorithm is showing that a sample set of polynomial size is sufficient to obtain a multiplicative approximation to the exact solution count.

The above is formalized in the following theorem.
\begin{theorem} \label{thm:jvv}
	(Adapted from Theorem 6.4 in~\cite{jerrumRandomGenerationCombinatorial1986} and Theorem 13.3 in~\cite{mooreNatureComputation2011}) If a self-reducible \textsf{\#P} problem on $n$ Boolean variables admits a solution generator with nonuniformity $\eta'$ such that $\eta' \leq \eta = O(\varepsilon n^{-1})$, then the solution count can be approximated, with high probability, to within a multiplicative factor of $1 + \varepsilon$ using only a polynomial number of calls to the generator.
\end{theorem}
Here, ``with high probability'' is in the sense of a polynomial-time randomized approximation scheme (PRAS), which implies a confidence $1-\delta$ that the outcome is correct (a typical choice is $\delta=1/4$).

By requiring additionally a fully polynomial almost uniform generator, that is, a solution generator whose nonuniformity is at most $\eta$ and whose runtime is polynomial in $n$ and $\log (\eta^{-1})$, JVV provides a fully polynomial-time randomized approximation scheme (FPRAS) for approximate solution counting of self-reducible formulas. That is, for a self-reducible formula over $n$ variables, the JVV algorithm returns, in time polynomial in $n$ and $\varepsilon^{-1}$, a rational number $C \in \mathbb{Q}$ such that
\begin{equation}
	\operatorname{Pr}\left[(1+\varepsilon)^{-1} N \leq C \leq (1+\varepsilon) N \right] \geq 1-\delta \,,
\end{equation}
with confidence $1-\delta=3/4$. The algorithm requires $O(n^2 / \varepsilon^2)$ calls to the generator~\cite{jerrumCountingSamplingIntegrating2003}. Using a bootstrapping method, the failure probability can be reduced arbitrarily at a runtime cost of $\log(1/\delta)$.

\section{Variational quantum counting}
\label{sec:vqcount}

The main innovation we introduce in this work is the use of the output state $\ket{\psi(\vec\beta,\vec\gamma)}$ of a QAOA circuit, expressed in Eq.~\eqref{eq:final-state}, as the generator component of the JVV self-reduction. The result is a variational quantum algorithm for approximate counting.

Although this algorithm can be implemented with different QAOA variants, we restrict the analysis in this section to GM-QAOA in order to establish theoretical guarantees on sample complexity. As mentioned in Sec.~\ref{sec:qaoa}, GM-QAOA ensures that all solution amplitudes in $\ket{\psi(\vec\beta,\vec\gamma)}$ are exactly equal. Thus, whenever the outcome of a measurement of the qubits in the computational basis is a solution, then it is sampled uniformly at random from the solution set. In terms of the nonuniformity defined in Eq.~\eqref{eq:nonuniformity}, we therefore have $\eta=0$. The critical figure of merit for the efficacy of sampling from $\ket{\psi(\vec\beta,\vec\gamma)}$ is hence the success rate of GM-QAOA, that is, the probability that the measurement outcome is a solution, which we can define as
\begin{equation}
	r = \mel{ \psi(\vec\beta,\vec\gamma) }{ \hat{\cal P}_{\cal G} }{ \psi(\vec\beta,\vec\gamma) } \,,\label{eq:successrate}
\end{equation}
where $\hat{\cal P}_{\cal G}$ is a projector to the solution space $\cal G$ (i.e., the ground space of the Ising model representing the problem) given by
\begin{equation}
	\hat{\cal P}_{\cal G} = \sum_{x\in \cal G} \ketbra{x}{x} \,.
\end{equation}
With this definition, the performance of different sampling-based algorithms for counting can be parameterized on equal footing using the sampling success rate $r$, the multiplicative error $\varepsilon$, and the confidence $1-\delta$. This characterization extends to classical algorithms employing a generator that fails to produce a solution at a rate of $1-r$.

The output state $\ket{\psi(\vec{\beta}, \vec{\gamma})}$ of an optimized GM-QAOA circuit, combined with postselection to eliminate non-solutions, can be used straightforwardly as a solution generator to a self-reducible problem. However, the JVV algorithm requires also a solver for each subproblem. Direct use of the GM-QAOA circuit optimized for the full problem would require additional postselection on the variable fixed in the subproblem.

\begin{figure}[t!]
	\subfloat[]{
		\includegraphics[width=0.9\columnwidth]{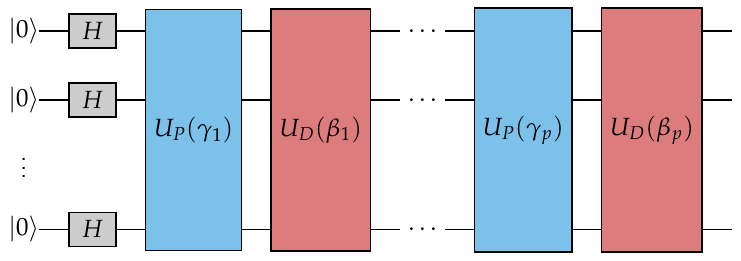}
	} \\
	\subfloat[]{
		\includegraphics[width=0.9\columnwidth]{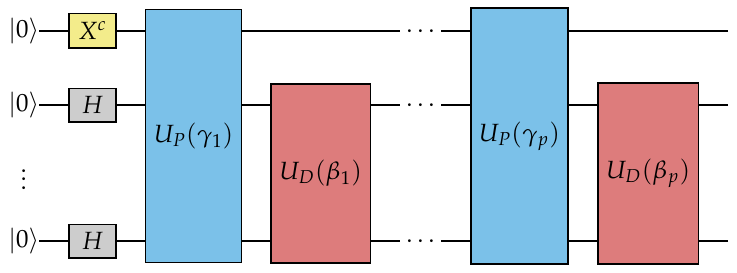}
	}
	\caption{QAOA solution sampler before (a) and after (b) fixing the first qubit to $c$ during the self-reduction procedure. The Hadamard gate $H$ acting on the fixed qubit is replaced by the Pauli-$X$ gate, conditioned classically on $c$. For each layer, the problem operator $U_{P}(\gamma)$ remains unchanged, while the gates of drive operator $U_{D}(\beta)$ acting on the fixed qubit are removed.}
	\label{fig:qaoa-self-reduction}
\end{figure}

We address this bottleneck by modifying the GM-QAOA circuit for each subproblem as illustrated in Fig.~\ref{fig:qaoa-self-reduction}. For each variable fixed during the self-reduction, we replace the initial Hadamard gate with a $X^c$ operation, where $c=0,1$ according to whether we are fixing the qubit to 0 or 1, and remove the mixer gates that act on the corresponding qubit. This yields a reduced GM-QAOA circuit for the subproblem with a variable fixed. The optimal parameters of the reduced circuit need not be the same as those of the original circuit, so in principle additional optimization may be required to reach optimality for the reduced circuit. On the other hand, it is interesting to ask whether the success rate of the reduced circuit is equal to or higher than that of the original circuit \emph{with the same parameters}, since in this case we may choose to skip the reoptimization for the reduced circuits.

The answer to the above question is affirmative in the case of shallow GM-QAOA circuits in the limit $\beta_i = \gamma_i = \pi, i = 1, \dots, p$. In this limit, GM-QAOA reduces to the Grover algorithm. To see this, note that for any $H_P$ that can be expressed as sums of Pauli strings, we can write a circuit that collapses all excited levels of $H_P$ to a single excited level, leading to a two-level Hamiltonian evolution circuit $U_P$. With the aforementioned choice of parameters, $U_P$ then becomes an oracle for the ground states of $H_P$ (up to a constant phase) and $U_D$ becomes the Grover diffusion operator. In this limit, at a fixed depth $p=O(1)$, the success rate is known to be a non-decreasing function of the ratio of solutions to total number of states. If, as we descend the tree in Fig.~\ref{fig:jvv-algorithm}, we pick the branches that yield probabilities $\ge 1/2$, the ratio of solutions to total number of states is non-decreasing. This means that, for this choice of parameters, a shallow GM-QAOA circuit for a given problem can be modified as in Fig.~\ref{fig:qaoa-self-reduction} to solve subproblems with equal or greater success rate.

The above argument, which is exact only in the Grover limit, suggests that the strategy of Fig.~\ref{fig:qaoa-self-reduction} may also hold away from that limit for GM-QAOA circuits. In the following section, we present numerical evidence to this effect and also compare the performance of regular QAOA circuits on the same task, for which the Grover limit is not relevant.

Guided by the intuition outlined above, we introduce a variational quantum algorithm for approximate counting called VQCount. Its steps are summarized in Algorithm~\ref{alg:vqcount} below. Given a self-reducible SAT formula $\varphi(x)$, we define the problem Hamiltonian $H_{P}$ whose ground states represent the solutions to $\varphi(x)$. In this work, we map $\varphi(x)$ to the Ising model using standard techniques~\cite{lucasIsingFormulationsMany2014} (see Sec.~\ref{sec:solving-sharp-P-complete-problems-with-vqcount} for details). A GM-QAOA circuit is optimized to minimize the expectation value of $H_{P}$. We then sample solutions by measuring the output of the circuit and postselecting solutions only. By virtue of the GM-QAOA algorithm, solutions are sampled uniformly. Once a sufficient number of samples is collected, we evaluate the conditional probabilities for the two values of a given variable and choose the value corresponding to the largest probability, as in the JVV algorithm. This choice leads to a reduced GM-QAOA circuit, as in Fig.~\ref{fig:qaoa-self-reduction}, and the process restarts with this circuit (but forgoing parameter optimization), until all variables are fixed.

\begin{algorithm}[t!]
	\caption{VQCount}\label{alg:vqcount}
	\begin{algorithmic}[1]
		\REQUIRE Number of variables: $n$, Problem Hamiltonian: $H_P$, Mixer Hamiltonian: $H_D$, Initial parameters: $(\vec{\beta}_0, \vec{\gamma}_0)$, Depth: $p$, Number of optimization steps: $n_{o}$, Number of solutions to sample: $n_s$
		\STATE PQC($\vec{\beta}_0, \vec{\gamma}_0) \leftarrow \text{QAOA-Circuit}(H_P, H_D, \vec{\beta}_0, \vec{\gamma}_0, p$)
			\STATE $\text{PQC}(\vec{\beta}, \vec{\gamma}) \leftarrow \text{Optimize}(\text{PQC}(\vec{\beta}_0, \vec{\gamma}_0), n_{o})$
			\STATE $\tilde{N} \leftarrow 1$
			\FOR{$i \in \{1, \dots, n\}$}
			\STATE $S \leftarrow \{ \ \}$
			\WHILE{$\abs{S} < n_{s}$}
			\STATE $m \leftarrow \text{Measure}(\text{PQC}(\vec{\beta}, \vec{\gamma}))$
			\IF{$\text{Verify-Solution}(m, H_{P})$}
			\STATE $S \leftarrow S \cup \{m\}$
			\ENDIF
			\ENDWHILE
			\STATE $w, \tilde{p} \leftarrow \text{Majority-Prefix}(S)$
			\STATE $\text{PQC}(\vec{\beta}, \vec{\gamma}) \leftarrow \text{Self-Reduce}(\text{PQC}(\vec{\beta}, \vec{\gamma}), w)$
			\STATE $\tilde{N} \leftarrow \tilde{N} / \tilde{p}$
			\ENDFOR
			\RETURN $\tilde{N}$
	\end{algorithmic}
\end{algorithm}

Since VQCount is built on the JVV algorithm (as described in Sec.~\ref{sec:jvv-algorithm}), its usage requires $O\left(\frac{n^2 \log(1/\delta)}{r \varepsilon^2}\right)$ samples to approximate the ground state count to multiplicative accuracy $\varepsilon$ with confidence $1-\delta$, given success rate $r$ of the GM-QAOA circuit, as defined in Eq.~\eqref{eq:successrate}. This compares favorably to related work on variational algorithms for counting weighted ground states of classical spin Hamiltonians~\cite{sundarQuantumAlgorithmCount2019}. When all configurations are assigned equal weight, the weighted counting problem reduces to the unweighted case, which requires $O\left(\frac{\sqrt{N \log (1/\delta)}}{r \varepsilon}\right)$ samples for the same task. Consequently, when $N=O(2^n)$, VQCount requires exponentially fewer samples than the methods of Ref.~\cite{sundarQuantumAlgorithmCount2019}. Alternative VQA circuits that implement solution sampling may also be used inside VQCount to obtain heuristics to which the theoretical sample-complexity guarantees derived above may not apply. We will investigate such variants in the next section.

A few comments are in order. First, generating quantum states that correspond to arbitrary probability distributions, including uniform distributions over combinatorial structures, is hard in general, and easy only in exceptional cases~\cite{groverCreatingSuperpositionsThat2002, aharonovAdiabaticQuantumState2003}. Furthermore, in general, GM-QAOA requires circuit depths scaling exponentially with problem size to achieve finite success rates~\cite{bridiAnalyticalResultsQuantum2024, xiePerformanceUpperBound2025}. These observations simply confirm that there is no ``free lunch'' for quantum counting algorithms: \textsf{\#P}-hard problems remain hard for quantum computers, assuming complexity theory hypotheses widely believed to be true. In light of this, one may ask instead whether hybrid algorithms perform well on average in practice. In the next section, we tackle this question for VQCount.

\section{Solving \textsf{\#P}-Complete Problems with VQCount}
\label{sec:solving-sharp-P-complete-problems-with-vqcount}

\begin{figure}[t!]
	\centering
	\includegraphics[width=0.9\columnwidth]{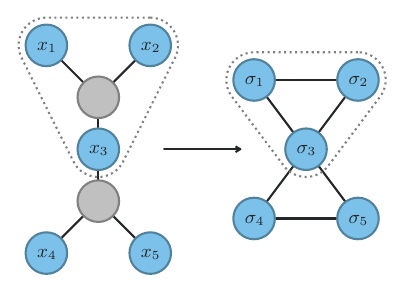}
	\caption{Example of the mapping to the Ising model for NAE3SAT / 1-in-3SAT formulae. Left panel: factor graph of the formula with variable (blue) and clause (grey) vertices. Right: Ising model corresponding to the formula represented in the left panel. For 1-in-3SAT, a magnetic field term is added to favor configurations in which each clause contains exactly one variable set to 1.}
	\label{fig:ising-mapping}
\end{figure}

In this section, we evaluate VQCount as a practical heuristic for solving \textsf{\#P}-complete counting problems. We deviate from the Grover limit of GM-QAOA introduced in the previous section in a few ways. First, we operate in the variational regime, where parameters $(\vec{\beta}, \vec{\gamma})$ are optimized. We also work with common problem Hamiltonians $H_P$ (see Eq.~\eqref{eq:Ising}) that contain towers of excited states, instead of the two-level oracle setting. Moreover, we are interested in studying the effects of nonuniformity in the output state distribution. To do this, we will replace the GM-QAOA primitive with a regular QAOA approach, namely, quantum approximate optimization. This amounts to passing $H_D^X$ as an input to Algorithm~\ref{alg:vqcount} instead of $H_D^{GM}$. Since quantum annealing does not sample degenerate ground states uniformly in general~\cite{matsudaGroundstateStatisticsAnnealing2009, mandraExponentiallyBiasedGroundState2017,goldenFairSamplingError2022}, we do not expect quantum approximate optimization to do so either.

\begin{figure*}[t!]
	\centering
	\includegraphics[width=0.85\textwidth]{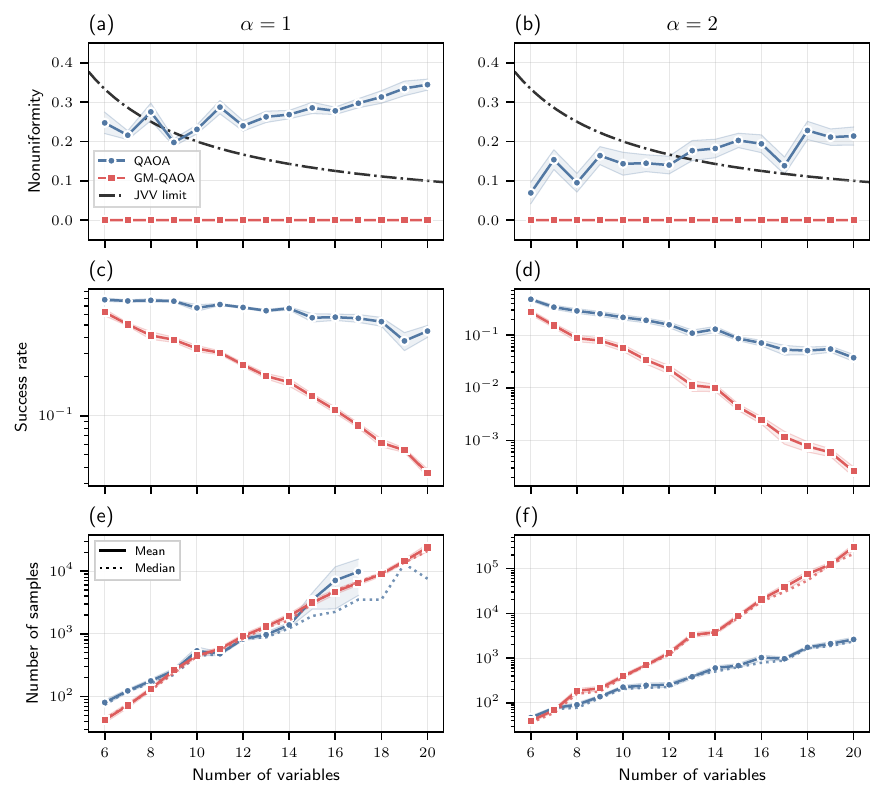}
	\caption{VQCount performance with depth $p=3$ QAOA (blue circles) and GM-QAOA (red squares) circuits for NAE3SAT instances at $\alpha=1$ (left panels) and $\alpha=2$ (right panels). Solid and dotted lines represent the mean and median, respectively. Shaded regions show the standard error of the mean. (a,b) Maximal nonuniformity throughout the self-reduction. The JVV algorithm requires the nonuniformity to be at most $O(n^{-1})$ for $n$ variables, as illustrated (dotted-dashed black lines) with the function $f(n) \propto 1/n$. (c,d) Minimal success rate throughout the self-reduction. (e,f) Numbers of samples needed to achieve an approximate count with error tolerance $\varepsilon = 1/3$. An exponential fit extrapolated from the last five points of the median indicates an asymptotic scaling of $(1.5 \pm 0.2)^n$ (QAOA) and $(1.44 \pm 0.02)^{n}$ (GM-QAOA) for (e), and $(1.34 \pm 0.06)^n$ (QAOA) and $(1.88 \pm 0.06 )^n$ (GM-QAOA) for (f).}
	\label{fig:nae3sat-tvd}
\end{figure*}

The counting problems we study are positive Not-All-Equal 3SAT (\#NAE3SAT) and positive 1-in-3SAT (\#1-in-3SAT). Both problems map to an Ising model partition function via the transformation $x_i \to \sigma_i = 2x_i - 1$~\cite{lucasIsingFormulationsMany2014}. First, we represent each problem instance as a bipartite graph, called a factor graph, in which variable vertices are connected to clause vertices, as illustrated in Fig.~\ref{fig:ising-mapping}. Each clause vertex is then replaced by a frustrated antiferromagnetic triangle in the Ising graph $G=(V,E)$, where $V$ is the vertex set representing spins and $E$ is the edge set representing interactions. The mapping of 1-in-3SAT to the Ising model additionally includes a longitudinal field term on each variable that imposes the 1-in-3 condition.

Concretely, the Ising Hamiltonian associated with these problems is
\begin{equation}
	H_P = - \sum_{(i,j) \in E} J_{ij} \sigma_i \sigma_j - \sum_{i \in V} h_i \sigma_i \,. \label{eq:Ising}
\end{equation}
Setting $J_{ij} = -1$ and $h_i = 0$ yields an exact mapping from ground states of $H_P$ to solutions of a NAE3SAT formula. For 1-in-3SAT, we add a longitudinal field $h_i = 1$.

Many random SAT formula families exhibit a phase transition phenomenon as a function of the density of clauses $\alpha = m/n$, where $m$ and $n$ denote the numbers of clauses and variables, respectively. The hardness of random \#NAE3SAT is maximal close to the satisfiability threshold, located at a critical clause-to-variable ratio $\alpha_c \approx 2.1$~\cite{achlioptasPhaseTransition1ink2001}. To study the effect of average formula hardness on VQCount performance, we generate instances at $\alpha = 1$ and $\alpha = 2$. We also restrict the sampled instances to connected biregular graphs. Similarly, positive $\#$1-in-3SAT instances are hardest close to the satisfiability threshold $\alpha_c \approx 2/3$~\cite{raymondPhaseDiagram1in32007}. To study this problem in the hard regime, we generate random cubic graphs, placing a clause on each vertex and a variable on each edge. $\#$1-in-3SAT close to the threshold belongs to the category of locked problems~\cite{zdeborovaStatisticalPhysicsHard2008}, in which the Hamming distance between any two solutions is $O(n)$. \textsf{\#P} problems in this regime are arguably the most challenging.

Tensor networks are a powerful tool for the simulation of quantum circuits. Shallow QAOA circuits, in particular, are amenable to tensor network simulation~\cite{grayHyperoptimizedTensorNetwork2021}. We use the library \textit{quimb}~\cite{grayQuimbPythonPackage2018} to model the optimization and sampling of QAOA and GM-QAOA circuits. Circuit parameters are optimized using sequential least squares programming, as implemented in the library \textit{SciPy}~\cite{virtanenSciPy10Fundamental2020}. To benchmark our results, we use the probabilistic exact model counter \textit{Ganak}~\cite{sharmaGANAKScalableProbabilistic2019} to find the exact number of solutions, and the SAT solver \textit{Glucose4}~\cite{eenExtensibleSATsolver2004, audemardPredictingLearntClauses2009} implemented in the \textit{PySAT}~\cite{ignatievPySATPythonToolkit2018} toolkit is used to enumerate all solutions.

\begin{figure}[t!]
	\centering
	\includegraphics[width=0.9\columnwidth]{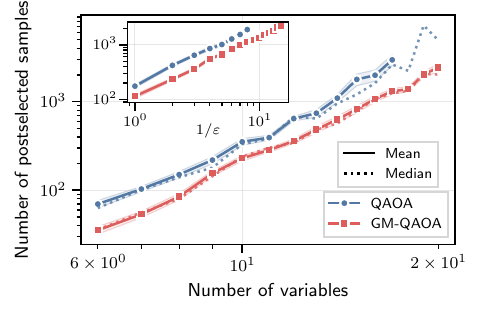}
	\caption{Number of post-selected samples needed to achieve an approximate count with error tolerance $\varepsilon=1/3$ with depth $p=3$ QAOA (blue circles) and GM-QAOA (red squares) circuits for NAE3SAT instances at $\alpha=1$. Solid and dotted lines represent the mean and median, respectively. Shaded regions show the standard error of the mean. A polynomial fit of the median indicates an asymptotic scaling of $n^{3.7 \pm 0.2}$ (QAOA) and $n^{3.42 \pm 0.07}$ (GM-QAOA). The inset panel shows the scaling behavior with the inverse of the error tolerance $\varepsilon$ for $n=12$ variables. A polynomial fit of the mean indicates an asymptotic scaling of $\varepsilon^{-1.10 \pm 0.03}$ (QAOA) and $\varepsilon^{-1.08 \pm 0.01}$ (GM-QAOA).}
	\label{fig:nae3sat-scaling}
\end{figure}

In the experiments presented below, we choose to sample solutions \textit{without} replacement, that is, we discard multiples. While this choice deviates from the JVV theorem prescription, it improves the efficacy of our heuristics whenever a problem has few solutions and / or the circuit output distribution is highly nonuniform.

\begin{figure}[t!]
	\centering
	\includegraphics[width=0.9\columnwidth]{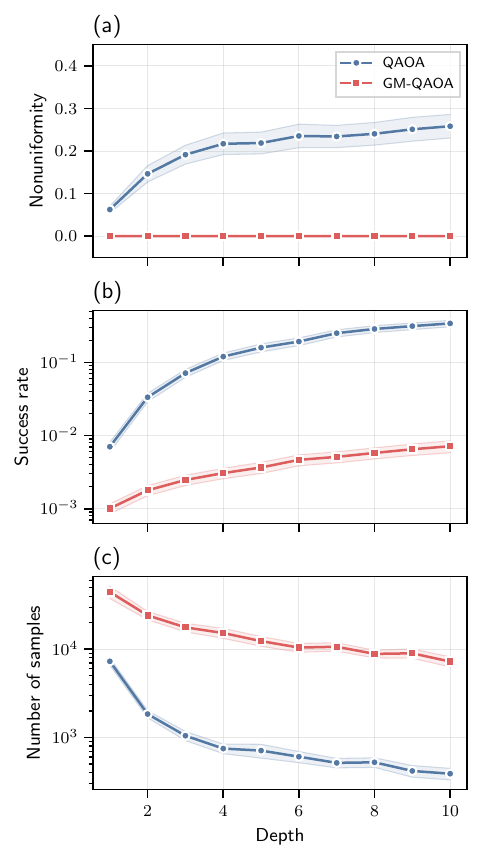}
	\caption{VQCount performance as a function of depth with QAOA (blue circles) and GM-QAOA (red squares) circuits for NAE3SAT instances of $n=16$ variables at $\alpha=2$. Solid lines represent the mean, and shaded regions show the standard error of the mean. (a) Maximal nonuniformity throughout the self-reduction. (b) Minimal success rate throughout the self-reduction. (c) Number of samples needed to achieve an approximate count with error tolerance error $\varepsilon = 1/3$.}
	\label{fig:nae3sat-depth}
\end{figure}

The performance of QAOA methods depends sensitively on the choice of $(\vec{\beta}_0,\vec{\gamma}_0)$. As the training of VQAs is \textsf{NP}-hard in general~\cite{bittelTrainingVariationalQuantum2021}, a good choice of initial parameters is necessary to avoid optimization pitfalls, such as rugged optimization landscapes and barren plateaus. One method of initialization is trotterized quantum annealing~\cite{sackQuantumAnnealingInitialization2021}, which has been found to greatly improve QAOA performance compared to random initialization. We will be using this technique in our simulations.

\begin{figure*}[t!]
	\centering
	\includegraphics[width=0.85\textwidth]{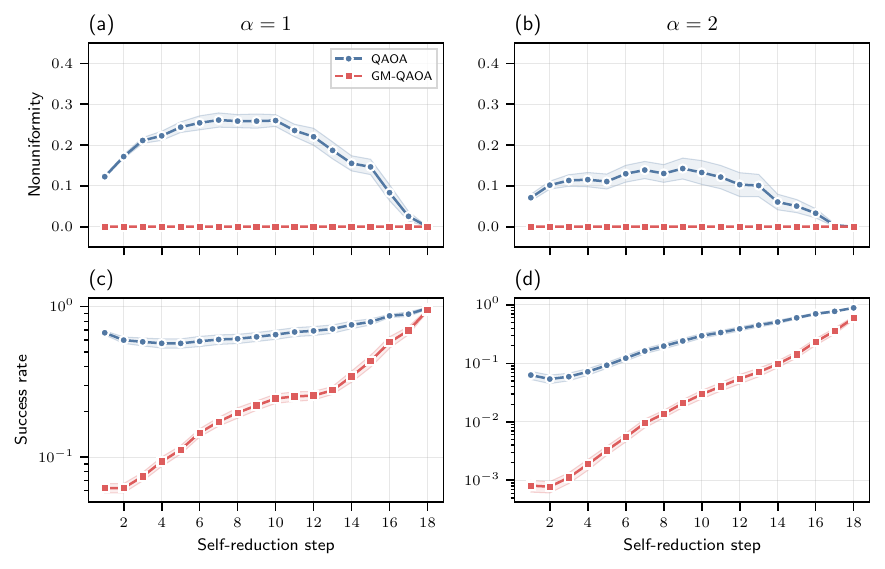}
	\caption{Impact of fixing qubits throughout the self-reduction on the sampling with depth $p=3$ QAOA (blue circles) and GM-QAOA (red squares) circuits for NAE3SAT instances of $n=18$ variables at $\alpha=1$ (left panels) and $\alpha=2$ (right panels). Solid lines represent the mean, and shaded regions show the standard error of the mean. (a,b) Nonuniformity as a function of the number of steps in the self-reduction. (c,d) Success rate as a function of the number of steps in the self-reduction.}
	\label{fig:nae3sat-self-reduction-step}
\end{figure*}

We begin by characterizing the performance of VQCount on \#NAE3SAT. For two clause densities $\alpha=1,2$, we generate 20 random instances per instance size, ranging from 6 to 20 variables. Fig.~\ref{fig:nae3sat-tvd} summarizes our results for circuits of depth $p=3$. Figs.~\ref{fig:nae3sat-tvd}(a) and~\ref{fig:nae3sat-tvd}(b) confirm that GM-QAOA samples solutions perfectly uniformly. In contrast, the distribution of solutions in the QAOA output is increasingly nonuniform with instance size.

Figs.~\ref{fig:nae3sat-tvd}(c) and~\ref{fig:nae3sat-tvd}(d) show that, at fixed depth, the success rate of GM-QAOA and QAOA circuits decreases roughly exponentially with instance size, though the decrease is faster for GM-QAOA. Success rates are lower and decay faster for $\alpha=2$ compared to $\alpha=1$, reflecting the different complexity of the problem close to and away from the satisfiability threshold.

Figs.~\ref{fig:nae3sat-tvd}(e) and~\ref{fig:nae3sat-tvd}(f) show the total number of samples (including those discarded by postselection) needed for VQCount to yield an approximate count within a multiplicative error of $\varepsilon = 1/3$. The procedure we use to determine this number is described in Appendix~\ref{app:min-samples}. The number of samples grows exponentially with instance size due to the postselection of solutions for both QAOA and GM-QAOA as backends for sampling. Despite the exponentially lower success rate of GM-QAOA, VQCount requires roughly equal numbers of samples with either QAOA or GM-QAOA at $\alpha=1$. In contrast, at $\alpha=2$, VQCount with QAOA requires exponentially fewer samples than with GM-QAOA, despite the nonuniformity in solution sampling with QAOA. This suggests that QAOA may be a more efficient sampling backend for VQCount in cases where solutions are few and far apart. We further investigate this behavior in our discussion of \#1-in-3SAT below. The number of post-selected samples (excluding those discarded by postselection) required for VQCount to produce an estimate within a given factor $\varepsilon$ of the exact count, presented in Fig.~\ref{fig:nae3sat-scaling}, appears polynomial in the number of variables and $1/\varepsilon$ as expected.

So far, we have investigated the performance of VQCount using QAOA circuits of fixed depth. Fig.~\ref{fig:nae3sat-depth} shows that, as the depth of the quantum circuit increases, the success rate increases, leading to a decrease in the number of samples required to achieve an approximate count within a given error. QAOA seems to improve faster with depth than GM-QAOA. On the other hand, QAOA nonuniformity increases and then plateaus with increasing depth.

In this work, the angle parameters of the QAOA circuit are only optimized once and held constant throughout the self-reduction process. In Sec.~\ref{sec:vqcount} we showed that, in the Grover limit, GM-QAOA success rates are non-decreasing with self-reduction step. We now test whether this is also the case away from the Grover limit for GM-QAOA and for regular QAOA in general. Our results are summarized in Fig.~\ref{fig:nae3sat-self-reduction-step}. We observe that success rates remain non-decreasing for GM-QAOA. In contrast, in QAOA nonuniformity initially increases before decreasing, as seen in Fig.~\ref{fig:nae3sat-self-reduction-step}(a,b), with the increase being more pronounced for $\alpha=1$. This effect correlates with a decrease in success rate in the first few self-reductions. The success rate of QAOA is nevertheless always superior to that of GM-QAOA for the instances we studied. A secondary reason for our choice not to re-optimize parameters is that re-optimization would compound the considerable computational cost of the tensor network simulations required to produce Fig.~\ref{fig:nae3sat-self-reduction-step}. A study of parameter re-optimization would be more amenable to VQCount runs on quantum hardware, which we hope to perform in future work.

The results shown thus far suggest that low-depth QAOA is a better heuristic solution sampler than GM-QAOA at the same depth for the problems studied here. GM-QAOA also depends on a $n$-qubit control gate, whose implementation poses a challenge for both near-term noisy quantum processors and classical simulation. In the last part of this section, we further study the performance of VQCount with the QAOA sampling backend as an approximate counting heuristic. To this end, we apply VQCount to positive \#1-in-3SAT at $\alpha=2/3$. We use known techniques to sample problem instances from the cubic graph ensemble described above.

\begin{figure}[t!]
	\centering
	\includegraphics[width=0.9\columnwidth]{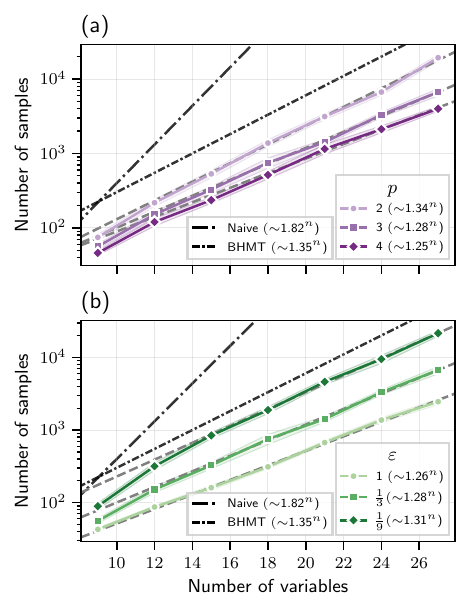}
	\caption{Scaling behavior of VQCount with QAOA circuits for 1-in-3SAT instances at $\alpha=2/3$. The number of samples needed to achieve an approximate count using naive rejection sampling and the BHMT quantum counting algorithm are shown with a black dash-dotted line and a black densely dash-dotted line, respectively, following the functions $f(n) \propto 1.82^{n}$ and $f(n) \propto 1.35^{n}$ for $n$ variables. Solid lines represent the mean, and shaded regions show the standard error of the mean. Dashed lines indicate the exponential fit extrapolated from the last four points as shown in the legend. (a) Number of samples needed for the approximate count to be within error tolerance $\varepsilon = 1/3$ for depths $p$. The exponential fit indicates an asymptotic scaling of $(1.34 \pm 0.02)^{n}$, $(1.28 \pm 0.01)^{n}$, and $(1.25 \pm 0.01)^{n}$. (b) Number of samples needed for the approximate count to be within error tolerance $\varepsilon$ for a depth $p=3$. The exponential fit indicates an asymptotic scaling of $(1.26 \pm 0.01)^{n}$, $(1.28 \pm 0.01)^{n}$, and $(1.31 \pm 0.01)^{n}$.}
	\label{fig:1in3sat-number-of-samples}
\end{figure}

\begin{figure}[t!]
	\centering
	\includegraphics[width=0.9\columnwidth]{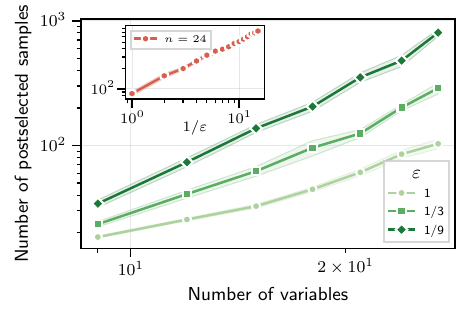}
	\caption{Number of post-selected samples needed to achieve an approximate count with error tolerance $\varepsilon$ with depth $p=3$ QAOA circuits for 1-in-3SAT instances at $\alpha=2/3$. The inset panel shows the scaling behavior with the inverse of the error tolerance $\varepsilon$ for $n=24$ variables, for which a polynomial fit indicates an asymptotic scaling of $\varepsilon^{-0.78 \pm 0.01}$. Solid lines represent the mean, and shaded regions show the standard error of the mean.}
	\label{fig:1in3sat-scaling}
\end{figure}

Fig.~\ref{fig:1in3sat-number-of-samples} summarizes the scaling behaviour of VQCount with low-depth QAOA. As seen in Fig.~\ref{fig:1in3sat-number-of-samples}(a), VQCount scales exponentially better than naive rejection sampling and its performance improves with circuit depth. Moreover, its performance is comparable to that of the quantum counting algorithm by Brassard, H\o yer, Mosca and Tapp (BHMT)~\cite{brassardQuantumAmplitudeAmplification2002}, which scales as $O(\sqrt{\frac{N}{M}})$, where $M$ is the size of the solution space and $N$ is the total size of the search space. At the depths we can access using reasonable computational resources, VQCount scales worse than state-of-the-art classical exact solvers for this problem~\cite{kourtisFastCountingTensor2019,dudekEfficientContractionLarge2020,dudekParallelWeightedModel2021}. The accuracy of VQCount can be improved by increasing the number of samples taken, as shown in Fig.~\ref{fig:1in3sat-number-of-samples}(b). Finally, Fig.~\ref{fig:1in3sat-scaling} shows the number of post-selected samples needed to produce an estimate within a given factor $\varepsilon$ of the exact count as a function of number of variables and $1/\varepsilon$. Our results suggest a superpolynomially but subexponentially increasing number of post-selected samples (cf. Fig.~\ref{fig:nae3sat-scaling} for \#NAE3SAT).

Classical sampling-based approximate counting methods struggle to generate samples almost uniformly from complex solution spaces, prompting the use of hashing-based counting algorithms. VQCount proposes a tunable tradeoff between classical and quantum resources (postselection cost vs. ansatz depth) for addressing the solution sampling challenge.

\section{Conclusion}

This work introduces VQCount, a sampling-based variational quantum heuristic for approximate model counting based on QAOA and the JVV theorem. We establish the soundness of VQCount in a limit where it is shown to reproduce Grover's algorithm. We show that, compared to previous work~\cite{sundarQuantumAlgorithmCount2019}, VQCount achieves an exponential improvement in the number of samples required for obtaining an estimate within an arbitrarily small multiplicative factor of the exact count whenever the number of solutions is exponentially large.

We then numerically investigate the performance of VQCount, with both GM-QAOA and regular QAOA as sampling subroutines, as a heuristic (without theoretical guarantees), using tensor network contraction to simulate the quantum circuit component of the algorithm. For the counting problems $\#$NAE3SAT and $\#$1-in-3SAT, we establish trade-offs between solution sampling success rate and sampling uniformity. These results establish VQCount as a viable framework for approximate counting using VQAs.

The implementation introduced here can serve as a baseline for future improvements in VQCount, including problem-aware mixers, warm-start strategies, and improved parameter optimization heuristics. While the current scaling of VQCount remains worse than that of modern classical solvers, Fig.~\ref{fig:1in3sat-number-of-samples}(a) indicates consistent performance improvement with increasing ansatz depth. This improvement leaves open the possibility that VQCount with circuits of (potentially variable) larger depth may eventually become competitive with classical heuristics for approximate counting.

Prior work has shown that accurate approximate counting is possible even with non-uniform sampling~\cite{gomesSamplingModelCounting2007}. It is plausible that similar arguments may also hold for VQCount. Finally, the only post-processing used in this work is postselection to reject samples outside the solution set. However, more elaborate post-processing, such as rejection-free cluster updates at zero temperature~\cite{ochoaFeedingMultitudePolynomialtime2019}, could be employed to reduce the postselection cost.

\section*{Acknowledgments}

We thank Jeremy Côté for critical reading of the manuscript. This work was supported by the Minist\`{e}re de l'\'{E}conomie et de l'Innovation du Qu\'{e}bec through a Research Chair in Quantum Computing, a Discovery grant awarded by the Natural Sciences and Engineering Research Council of Canada, and funding from the Canada First Research Excellence Fund.

\section*{Data availability statement}

The data that support the findings of this study are openly available at the following URL/DOI: \url{https://github.com/QuICoPhy-Lab/VQCount}~\cite{drapeauVQCount2025}.

\appendix

\section{Determining the minimum number of samples required to reach a given accuracy with VQCount}
\label{app:min-samples}

\begin{figure}[t!]
	\centering
	\includegraphics[width=0.9\columnwidth]{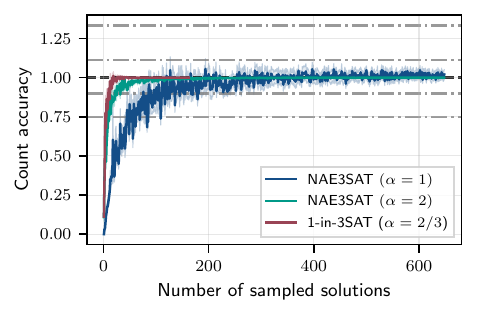}
	\caption{Count accuracy obtained by increasing the number of solutions sampled at each step of the self-reduction with depth $p=3$ QAOA circuits for NAE3SAT and 1-in-3SAT instances of $n=18$ variables. Solid lines represent the mean, and shaded regions show the standard error of the mean. The dashed black line represent the exact count accuracy, while the dash-dotted grey lines correspond to a multiplicative error of $\varepsilon = 1/3$ and $\varepsilon = 1/9$. }
	\label{fig:count-accuracy}
\end{figure}

To determine the minimum number of samples necessary to reach an approximate count with a multiplicative error $\varepsilon$, the VQCount algorithm was run by increasing the number of solutions sampled from the optimized QAOA circuit until the approximate number of solutions was inside the given error bounds. Note that the minimum number of samples includes those discarded by postselection. Fig.~\ref{fig:count-accuracy} shows the count accuracy, i.e., the approximate number of solutions over the exact number of solutions, for given error bounds. VQCount systematically underestimates the number of solutions, as seen most prominently in the $\#$NAE3SAT curve for $\alpha=1$.

\section{Sample efficiency of VQCount}

\begin{figure*}[t!]
	\centering
	\includegraphics[width=0.8\textwidth]{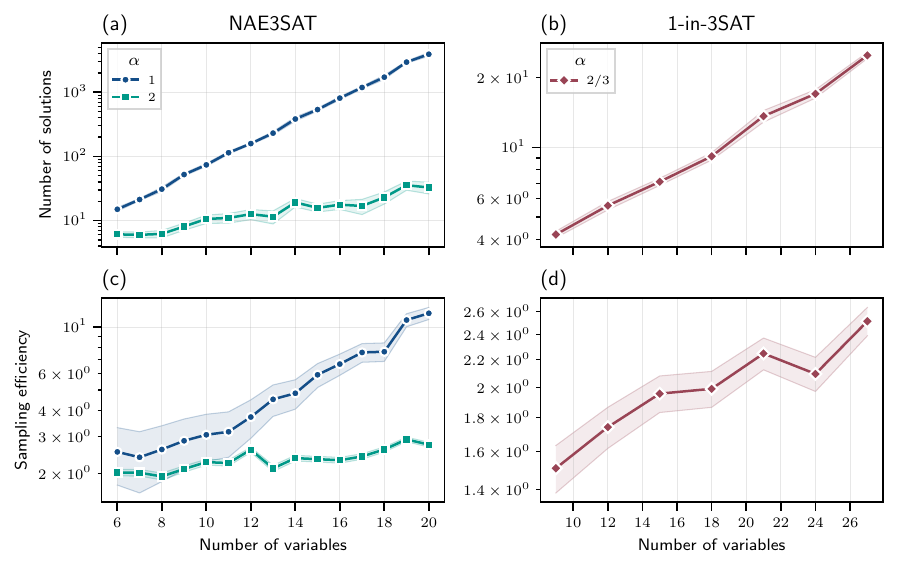}
	\caption{Number of solutions (a) and sampling efficiency (b) to achieve an error tolerance $\varepsilon = 1$ with depth $p=3$ QAOA circuits for NAE3SAT (left panels) and 1-in-3SAT (right panels) instances. Solid lines represent the mean, and shaded regions show the standard error of the mean.}
	\label{fig:sampling-efficiency}
\end{figure*}

The expectation with the VQCount algorithm is that only a polynomial number of samples is needed to count an exponential number of solutions. Since in numerical simulations we may encounter problem instances with few solutions, it is important to ascertain that VQCount needs fewer samples than the number of solutions in practice, to ensure that the algorithm is not naively sampling all solutions. Fig.~\ref{fig:sampling-efficiency} presents the sampling efficiency of VQCount, i.e., the exact number of solutions over the number of distinct solutions used in the algorithm, for the two problems studied in this work. We see that sampling efficiency correlates positively with solution density. Nevertheless, VQCount always uses fewer samples than the total number of solutions.

\newpage

\bibliography{references}

@inproceedings{aaronsonComputationalComplexityLinear2011,
  title = {The Computational Complexity of Linear Optics},
  booktitle = {Proceedings of the Forty-Third Annual {{ACM}} Symposium on {{Theory}} of Computing},
  author = {Aaronson, Scott and Arkhipov, Alex},
  year = {2011},
  month = jun,
  pages = {333--342},
  doi = {10.1145/1993636.1993682},
  isbn = {978-1-4503-0691-1},
}

@inproceedings{aaronsonQuantumApproximateCounting2020,
  title = {Quantum {{Approximate Counting}}, {{Simplified}}},
  booktitle = {Proceedings of the 2020 {{Symposium}} on {{Simplicity}} in {{Algorithms}}},
  author = {Aaronson, Scott and Rall, Patrick},
  year = {2020},
  pages = {24--32},
  doi = {10.1137/1.9781611976014.5},
}

@article{abramsonHailfinderBayesianSystem1996,
  title = {Hailfinder: {{A Bayesian}} System for Forecasting Severe Weather},
  shorttitle = {Hailfinder},
  author = {Abramson, Bruce and Brown, John and Edwards, Ward and Murphy, Allan and Winkler, Robert L.},
  year = {1996},
  month = mar,
  journal = {International Journal of Forecasting},
  volume = {12},
  number = {1},
  pages = {57--71},
  issn = {0169-2070},
  doi = {10.1016/0169-2070(95)00664-8},
}

@inproceedings{achlioptasPhaseTransition1ink2001,
  title = {The Phase Transition in 1-in-k {{SAT}} and {{NAE}} 3-{{SAT}}},
  booktitle = {Proceedings of the Twelfth Annual {{ACM-SIAM}} Symposium on {{Discrete}} Algorithms},
  author = {Achlioptas, Dimitris and Chtcherba, Arthur and Istrate, Gabriel and Moore, Cristopher},
  year = {2001},
  month = jan,
  pages = {721--722},
  isbn = {978-0-89871-490-6},
  nolink = {},
}

@inproceedings{aharonovAdiabaticQuantumState2003,
  title = {Adiabatic {{Quantum State Generation}} and {{Statistical Zero Knowledge}}},
  booktitle = {Proceedings of the {{Thirty-Fifth Annual ACM Symposium}} on {{Theory}} of {{Computing}}},
  author = {Aharonov, Dorit and {Ta-Shma}, Amnon},
  year = {2003},
  pages = {20--29},
  doi = {10.1145/780542.780546},
  isbn = {1-58113-674-9},
}

@inproceedings{audemardPredictingLearntClauses2009,
  title = {Predicting Learnt Clauses Quality in Modern {{SAT}} Solvers},
  booktitle = {Proceedings of the Twenty-First {{International Joint Conference}} on {{Artificial Intelligence}}},
  author = {Audemard, Gilles and Simon, Laurent},
  year = {2009},
  month = jul,
  pages = {399--404},
  nolink = {},
}

@inproceedings{balutaQuantitativeVerificationNeural2019,
  title = {Quantitative {{Verification}} of {{Neural Networks}} and {{Its Security Applications}}},
  booktitle = {Proceedings of the 2019 {{ACM SIGSAC Conference}} on {{Computer}} and {{Communications Security}}},
  author = {Baluta, Teodora and Shen, Shiqi and Shinde, Shweta and Meel, Kuldeep S. and Saxena, Prateek},
  year = {2019},
  month = nov,
  pages = {1249--1264},
  doi = {10.1145/3319535.3354245},
  isbn = {978-1-4503-6747-9},
}

@inproceedings{bartschiGroverMixersQAOA2020,
  title = {Grover {{Mixers}} for {{QAOA}}: {{Shifting Complexity}} from {{Mixer Design}} to {{State Preparation}}},
  shorttitle = {Grover {{Mixers}} for {{QAOA}}},
  booktitle = {Proceedings of the 2020 {{IEEE International Conference}} on {{Quantum Computing}} and {{Engineering}}},
  author = {B{\"a}rtschi, Andreas and Eidenbenz, Stephan},
  year = {2020},
  month = oct,
  pages = {72--82},
  doi = {10.1109/QCE49297.2020.00020},
}

@book{biereHandbookSatisfiability2021,
  title = {Handbook of {{Satisfiability}}},
  shorttitle = {Handbook of {{Satisfiability}}},
  author = {Biere, Armin and Heule, Marijn and {van Maaren}, Hans and Walsh, Toby},
  year = 2021,
  month = apr,
  edition = {2},
  volume = {336},
  publisher = {IOS Press},
  isbn = {978-1-64368-160-3},
}

@article{bittelTrainingVariationalQuantum2021,
  title = {Training {{Variational Quantum Algorithms Is NP-Hard}}},
  author = {Bittel, Lennart and Kliesch, Martin},
  year = {2021},
  month = sep,
  journal = {Physical Review Letters},
  volume = {127},
  number = {12},
  pages = {120502},
  publisher = {American Physical Society},
  doi = {10.1103/PhysRevLett.127.120502},
}

@article{boulandComplexityVerificationQuantum2019,
  title = {On the Complexity and Verification of Quantum Random Circuit Sampling},
  author = {Bouland, Adam and Fefferman, Bill and Nirkhe, Chinmay and Vazirani, Umesh},
  year = {2019},
  month = feb,
  journal = {Nature Physics},
  volume = {15},
  number = {2},
  pages = {159--163},
  publisher = {Nature Publishing Group},
  issn = {1745-2481},
  doi = {10.1038/s41567-018-0318-2},
}

@incollection{brassardQuantumAmplitudeAmplification2002,
  title = {Quantum Amplitude Amplification and Estimation},
  booktitle = {Quantum Computation and Information},
  author = {Brassard, Gilles and H{\o}yer, Peter and Mosca, Michele and Tapp, Alain},
  year = {2002},
  volume = {305},
  pages = {53--74},
  publisher = {American Mathematical Society},
  doi = {10.1090/conm/305/05215},
  isbn = {978-0-8218-2140-4},
  mrnumber = {1947332},
}

@article{bridiAnalyticalResultsQuantum2024,
  title = {Analytical Results for the Quantum Alternating Operator Ansatz with {{Grover}} Mixer},
  author = {Bridi, Guilherme Adamatti and Marquezino, Franklin de Lima},
  year = {2024},
  month = nov,
  journal = {Physical Review A},
  volume = {110},
  number = {5},
  pages = {052409},
  publisher = {American Physical Society},
  doi = {10.1103/PhysRevA.110.052409},
}

@inproceedings{broderHowHardIt1986,
  title = {How {{Hard}} Is {{It}} to {{Marry}} at {{Random}}? ({{On}} the {{Approximation}} of the {{Permanent}})},
  booktitle = {Proceedings of the {{Eighteenth Annual ACM Symposium}} on {{Theory}} of {{Computing}}},
  author = {Broder, Andrei Z.},
  year = {1986},
  pages = {50--58},
  doi = {10.1145/12130.12136},
  isbn = {0-89791-193-8},
}

@article{cerezoVariationalQuantumAlgorithms2021,
  title = {Variational Quantum Algorithms},
  author = {Cerezo, M. and Arrasmith, Andrew and Babbush, Ryan and Benjamin, Simon C. and Endo, Suguru and Fujii, Keisuke and McClean, Jarrod R. and Mitarai, Kosuke and Yuan, Xiao and Cincio, Lukasz and Coles, Patrick J.},
  year = {2021},
  month = sep,
  journal = {Nature Reviews Physics},
  volume = {3},
  number = {9},
  pages = {625--644},
  publisher = {Nature Publishing Group},
  issn = {2522-5820},
  doi = {10.1038/s42254-021-00348-9},
}

@inproceedings{chakrabortyScalableApproximateModel2013,
  title = {A {{Scalable Approximate Model Counter}}},
  booktitle = {Principles and {{Practice}} of {{Constraint Programming}}},
  author = {Chakraborty, Supratik and Meel, Kuldeep S. and Vardi, Moshe Y.},
  year = {2013},
  month = sep,
  pages = {200--216},
  doi = {10.1007/978-3-642-40627-0_18},
  isbn = {978-3-642-40627-0},
}

@misc{drapeauVQCount2025,
  title = {{VQCount}},
  author = {Drapeau, Julien},
  year = {2025},
  url = {https://github.com/QuICoPhy-Lab/VQCount},
}

@misc{dudekEfficientContractionLarge2020,
  title = {Efficient {{Contraction}} of {{Large Tensor Networks}} for {{Weighted Model Counting}} through {{Graph Decompositions}}},
  author = {Dudek, Jeffrey M. and {Due{\~n}as-Osorio}, Leonardo and Vardi, Moshe Y.},
  year = {2020},
  month = apr,
  number = {arXiv:1908.04381},
  eprint = {1908.04381},
  primaryclass = {cs},
  publisher = {arXiv},
  doi = {10.48550/arXiv.1908.04381},
  archiveprefix = {arXiv},
}

@misc{dudekParallelWeightedModel2021,
  title = {Parallel {{Weighted Model Counting}} with {{Tensor Networks}}},
  author = {Dudek, Jeffrey M. and Vardi, Moshe Y.},
  year = {2021},
  month = jun,
  number = {arXiv:2006.15512},
  eprint = {2006.15512},
  primaryclass = {cs},
  publisher = {arXiv},
  doi = {10.48550/arXiv.2006.15512},
  archiveprefix = {arXiv},
}

@inproceedings{duenas-osorioCountingbasedReliabilityEstimation2017,
  title = {Counting-Based Reliability Estimation for Power-Transmission Grids},
  booktitle = {Proceedings of the {{AAAI Conference}} on {{Artificial Intelligence}}},
  author = {{Duenas-Osorio}, Leonardo and Meel, Kuldeep and Paredes, Roger and Vardi, Moshe},
  year = {2017},
  month = feb,
  volume = {31},
  pages = {4488--4494},
  doi = {10.1609/aaai.v31i1.11178},
}

@inproceedings{eenExtensibleSATsolver2004,
  title = {An {{Extensible SAT-solver}}},
  booktitle = {Proceedings of the Sixth International Conference on Theory and {{Applications}} of {{Satisfiability Testing}}},
  author = {E{\'e}n, Niklas and S{\"o}rensson, Niklas},
  year = {2004},
  pages = {502--518},
  doi = {10.1007/978-3-540-24605-3_37},
  isbn = {978-3-540-24605-3},
}

@misc{farhiQuantumApproximateOptimization2014,
  title = {A {{Quantum Approximate Optimization Algorithm}}},
  author = {Farhi, Edward and Goldstone, Jeffrey and Gutmann, Sam},
  year = {2014},
  month = nov,
  number = {arXiv:1411.4028},
  eprint = {1411.4028},
  primaryclass = {quant-ph},
  publisher = {arXiv},
  doi = {10.48550/arXiv.1411.4028},
  archiveprefix = {arXiv},
}

@inproceedings{gomesSamplingModelCounting2007,
  title = {From Sampling to Model Counting},
  booktitle = {Proceedings of the Twentieth International Joint Conference on {{Artificial}} Intelligence},
  author = {Gomes, Carla P. and Hoffmann, Joerg and Sabharwal, Ashish and Selman, Bart},
  year = 2007,
  month = jan,
  pages = {2293--2299},
}

@article{goldenFairSamplingError2022,
  title = {Fair {{Sampling Error Analysis}} on {{NISQ Devices}}},
  author = {Golden, John and B{\"a}rtschi, Andreas and O'Malley, Daniel and Eidenbenz, Stephan},
  year = {2022},
  month = may,
  journal = {ACM Transactions on Quantum Computing},
  volume = {3},
  number = {2},
  pages = {8:1--8:23},
  doi = {10.1145/3510857},
}

@article{grayHyperoptimizedTensorNetwork2021,
  title = {Hyper-Optimized Tensor Network Contraction},
  author = {Gray, Johnnie and Kourtis, Stefanos},
  year = {2021},
  month = mar,
  journal = {Quantum},
  volume = {5},
  pages = {410},
  publisher = {Verein zur F{\"o}rderung des Open Access Publizierens in den Quantenwissenschaften},
  doi = {10.22331/q-2021-03-15-410},
}

@article{grayQuimbPythonPackage2018,
  title = {Quimb: {{A}} Python Package for Quantum Information and Many-Body Calculations},
  shorttitle = {Quimb},
  author = {Gray, Johnnie},
  year = {2018},
  month = sep,
  journal = {Journal of Open Source Software},
  volume = {3},
  number = {29},
  pages = {819},
  issn = {2475-9066},
  doi = {10.21105/joss.00819},
}

@misc{groverCreatingSuperpositionsThat2002,
  title = {Creating Superpositions That Correspond to Efficiently Integrable Probability Distributions},
  author = {Grover, Lov and Rudolph, Terry},
  year = {2002},
  month = aug,
  number = {arXiv:quant-ph/0208112},
  eprint = {quant-ph/0208112},
  publisher = {arXiv},
  doi = {10.48550/arXiv.quant-ph/0208112},
  archiveprefix = {arXiv},
}

@inproceedings{groverFastQuantumMechanical1996,
  title = {A {{Fast Quantum Mechanical Algorithm}} for {{Database Search}}},
  booktitle = {Proceedings of the {{Twenty-Eighth Annual ACM Symposium}} on {{Theory}} of {{Computing}}},
  author = {Grover, Lov K.},
  year = {1996},
  pages = {212--219},
  doi = {10.1145/237814.237866},
  isbn = {0-89791-785-5},
}

@inproceedings{hadfieldQuantumApproximateOptimization2017,
  title = {Quantum {{Approximate Optimization}} with {{Hard}} and {{Soft Constraints}}},
  booktitle = {Proceedings of the {{Second International Workshop}} on {{Post Moore's Era Supercomputing}}},
  author = {Hadfield, Stuart and Wang, Zhihui and Rieffel, Eleanor G. and O'Gorman, Bryan and Venturelli, Davide and Biswas, Rupak},
  year = {2017},
  month = nov,
  pages = {15--21},
  doi = {10.1145/3149526.3149530},
  isbn = {978-1-4503-5126-3},
}

@article{hadfieldQuantumApproximateOptimization2019,
  title = {From the {{Quantum Approximate Optimization Algorithm}} to a {{Quantum Alternating Operator Ansatz}}},
  author = {Hadfield, Stuart and Wang, Zhihui and O'Gorman, Bryan and Rieffel, Eleanor G. and Venturelli, Davide and Biswas, Rupak},
  year = {2019},
  month = feb,
  journal = {Algorithms},
  volume = {12},
  number = {2},
  pages = {34},
  publisher = {Multidisciplinary Digital Publishing Institute},
  issn = {1999-4893},
  doi = {10.3390/a12020034},
}

@incollection{hemaspaandraPowerSelfReducibilitySelectivity2020,
  title = {The {{Power}} of {{Self-Reducibility}}: {{Selectivity}}, {{Information}}, and {{Approximation}}},
  booktitle = {Complexity and {{Approximation}}: {{In Memory}} of {{Ker-I Ko}}},
  author = {Hemaspaandra, Lane A.},
  year = {2020},
  pages = {19--47},
  publisher = {Springer International Publishing},
  doi = {10.1007/978-3-030-41672-0_3},
  isbn = {978-3-030-41672-0},
}

@inproceedings{ignatievPySATPythonToolkit2018,
  title = {{{PySAT}}: {{A Python Toolkit}} for {{Prototyping}} with {{SAT Oracles}}},
  shorttitle = {{{PySAT}}},
  booktitle = {Proceedings of the Twenty-First International Conference on Theory and {{Applications}} of {{Satisfiability Testing}}},
  author = {Ignatiev, Alexey and Morgado, Antonio and {Marques-Silva}, Joao},
  year = {2018},
  pages = {428--437},
  doi = {10.1007/978-3-319-94144-8_26},
  isbn = {978-3-319-94144-8},
}

@book{jerrumCountingSamplingIntegrating2003,
  title = {Counting, {{Sampling}} and {{Integrating}}: {{Algorithm}} and {{Complexity}}},
  shorttitle = {Counting, {{Sampling}} and {{Integrating}}},
  author = {Jerrum, Mark},
  year = {2003},
  publisher = {Birkh{\"a}user},
  doi = {10.1007/978-3-0348-8005-3},
  isbn = {978-3-7643-6946-0 978-3-0348-8005-3},
}

@article{jerrumPolynomialtimeApproximationAlgorithms1993,
  title = {Polynomial-Time Approximation Algorithms for the {{Ising}} Model},
  author = {Jerrum, Mark and Sinclair, Alistair},
  year = {1993},
  month = oct,
  journal = {SIAM Journal on Computing},
  volume = {22},
  number = {5},
  pages = {1087--1116},
  publisher = {{Society for Industrial and Applied Mathematics}},
  issn = {0097-5397},
  doi = {10.1137/0222066},
}

@article{jerrumRandomGenerationCombinatorial1986,
  title = {Random Generation of Combinatorial Structures from a Uniform Distribution},
  author = {Jerrum, Mark R. and Valiant, Leslie G. and Vazirani, Vijay V.},
  year = {1986},
  month = jan,
  journal = {Theoretical Computer Science},
  volume = {43},
  pages = {169--188},
  issn = {0304-3975},
  doi = {10.1016/0304-3975(86)90174-X},
}

@article{kourtisFastCountingTensor2019,
  title = {Fast Counting with Tensor Networks},
  author = {Kourtis, Stefanos and Chamon, Claudio and Mucciolo, Eduardo and Ruckenstein, Andrei E.},
  year = {2019},
  month = nov,
  journal = {SciPost Physics},
  volume = {7},
  number = {5},
  pages = {060},
  issn = {2542-4653},
  doi = {10.21468/SciPostPhys.7.5.060},
}

@inproceedings{lagniezImprovedDecisionDNNFCompiler2017,
  title = {An Improved Decision-{{DNNF}} Compiler},
  booktitle = {Proceedings of the Twenty-Sixth International Joint Conference on Artificial Intelligence},
  author = {Lagniez, Jean-Marie and Marquis, Pierre},
  year = 2017,
  pages = {667--673},
  doi = {10.24963/ijcai.2017/93},
}

@article{lucasIsingFormulationsMany2014,
  title = {Ising Formulations of Many {{NP}} Problems},
  author = {Lucas, Andrew},
  year = {2014},
  journal = {Frontiers in Physics},
  volume = {2},
  pages = {5},
  issn = {2296-424X},
  doi = {10.3389/fphy.2014.00005},
}

@article{mandraExponentiallyBiasedGroundState2017,
  title = {Exponentially {{Biased Ground-State Sampling}} of {{Quantum Annealing Machines}} with {{Transverse-Field Driving Hamiltonians}}},
  author = {Mandr{\`a}, Salvatore and Zhu, Zheng and Katzgraber, Helmut G.},
  year = {2017},
  month = feb,
  journal = {Physical Review Letters},
  volume = {118},
  number = {7},
  pages = {070502},
  doi = {10.1103/PhysRevLett.118.070502},
}

@article{matsudaGroundstateStatisticsAnnealing2009,
  title = {Ground-State Statistics from Annealing Algorithms: Quantum versus Classical Approaches},
  shorttitle = {Ground-State Statistics from Annealing Algorithms},
  author = {Matsuda, Yoshiki and Nishimori, Hidetoshi and Katzgraber, Helmut G.},
  year = {2009},
  month = jul,
  journal = {New Journal of Physics},
  volume = {11},
  number = {7},
  pages = {073021},
  publisher = {IOP Publishing},
  issn = {1367-2630},
  doi = {10.1088/1367-2630/11/7/073021},
}

@book{mooreNatureComputation2011,
  title = {The {{Nature}} of {{Computation}}},
  author = {Moore, Cristopher and Mertens, Stephan},
  year = {2011},
  month = aug,
  publisher = {Oxford University Press},
  doi = {10.1093/acprof:oso/9780199233212.001.0001},
  isbn = {978-0-19-923321-2},
}

@book{nielsenQuantumComputationQuantum2011,
  title = {Quantum {{Computation}} and {{Quantum Information}}: 10th {{Anniversary Edition}}},
  shorttitle = {Quantum {{Computation}} and {{Quantum Information}}},
  author = {Nielsen, Michael A. and Chuang, Isaac L.},
  year = {2011},
  publisher = {Cambridge University Press},
  isbn = {978-1-107-00217-3},
  nolink = {},
}

@article{ochoaFeedingMultitudePolynomialtime2019,
  title = {Feeding the Multitude: {{A}} Polynomial-Time Algorithm to Improve Sampling},
  shorttitle = {Feeding the Multitude},
  author = {Ochoa, Andrew J. and Jacob, Darryl C. and Mandr{\`a}, Salvatore and Katzgraber, Helmut G.},
  year = {2019},
  month = apr,
  journal = {Physical Review E},
  volume = {99},
  number = {4},
  pages = {043306},
  publisher = {American Physical Society},
  doi = {10.1103/PhysRevE.99.043306},
}

@article{raymondPhaseDiagram1in32007,
  title = {Phase Diagram of the 1-in-3 Satisfiability Problem},
  author = {Raymond, Jack and Sportiello, Andrea and Zdeborov{\'a}, Lenka},
  year = {2007},
  month = jul,
  journal = {Physical Review E},
  volume = {76},
  number = {1},
  pages = {011101},
  publisher = {American Physical Society},
  doi = {10.1103/PhysRevE.76.011101},
}

@article{rothHardnessApproximateReasoning1996,
  title = {On the Hardness of Approximate Reasoning},
  author = {Roth, Dan},
  year = {1996},
  month = apr,
  journal = {Artificial Intelligence},
  volume = {82},
  number = {1},
  pages = {273--302},
  issn = {0004-3702},
  doi = {10.1016/0004-3702(94)00092-1},
}

@article{sackQuantumAnnealingInitialization2021,
  title = {Quantum Annealing Initialization of the Quantum Approximate Optimization Algorithm},
  author = {Sack, Stefan H. and Serbyn, Maksym},
  year = {2021},
  month = jul,
  journal = {Quantum},
  volume = {5},
  pages = {491},
  publisher = {Verein zur F{\"o}rderung des Open Access Publizierens in den Quantenwissenschaften},
  issn = {2521-327X},
  doi = {10.22331/q-2021-07-01-491},
}

@inproceedings{sangCombiningComponentCaching2004,
  title = {Combining {{Component Caching}} and {{Clause Learning}} for {{Effective Model Counting}}},
  booktitle = {Proceedings of the Seventh International Conference on Theory and Applications of Satisfiability Testing},
  author = {Sang, Tian and Bacchus, Fahiem and Beame, Paul and Kautz, Henry. and Pitassi, Toniann},
  year = 2004,
  volume = {4},
}

@inproceedings{sangPerformingBayesianInference2005,
  title = {Performing {{Bayesian}} Inference by Weighted Model Counting},
  booktitle = {Proceedings of the Twentieth National Conference on {{Artificial}} Intelligence},
  author = {Sang, Tian and Bearne, Paul and Kautz, Henry},
  year = {2005},
  month = jul,
  volume = {1},
  pages = {475--481},
  isbn = {978-1-57735-236-5},
  nolink = {},
}

@inproceedings{sharmaGANAKScalableProbabilistic2019,
  title = {{{GANAK}}: {{A Scalable Probabilistic Exact Model Counter}}},
  shorttitle = {{{GANAK}}},
  booktitle = {Proceedings of the {{Twenty-Eighth International Joint Conference}} on {{Artificial Intelligence}}},
  author = {Sharma, Shubham and Roy, Subhajit and Soos, Mate and Meel, Kuldeep S.},
  year = {2019},
  month = jul,
  pages = {1169--1176},
  doi = {10.24963/ijcai.2019/163},
}

@article{sinclairApproximateCountingUniform1989,
  title = {Approximate Counting, Uniform Generation and Rapidly Mixing {{Markov}} Chains},
  author = {Sinclair, Alistair and Jerrum, Mark},
  year = {1989},
  journal = {Information and Computation},
  volume = {82},
  number = {1},
  pages = {93--133},
  issn = {0890-5401},
  doi = {10.1016/0890-5401(89)90067-9},
}

@inproceedings{stockmeyerComplexityApproximateCounting1983,
  title = {The Complexity of Approximate Counting},
  booktitle = {Proceedings of the Fifteenth Annual {{ACM}} Symposium on {{Theory}} of Computing},
  author = {Stockmeyer, Larry},
  year = 1983,
  month = dec,
  pages = {118--126},
  doi = {10.1145/800061.808740},
  isbn = {978-0-89791-099-6},
}

@misc{sundarQuantumAlgorithmCount2019,
  title = {A Quantum Algorithm to Count Weighted Ground States of Classical Spin {{Hamiltonians}}},
  author = {Sundar, Bhuvanesh and Paredes, Roger and Damanik, David T. and {Due{\~n}as-Osorio}, Leonardo and Hazzard, Kaden R. A.},
  year = {2019},
  month = aug,
  number = {arXiv:1908.01745},
  eprint = {1908.01745},
  primaryclass = {quant-ph},
  publisher = {arXiv},
  doi = {10.48550/arXiv.1908.01745},
  archiveprefix = {arXiv},
}

@article{valiantComplexityEnumerationReliability1979,
  title = {The {{Complexity}} of {{Enumeration}} and {{Reliability Problems}}},
  author = {Valiant, Leslie G.},
  year = {1979},
  month = aug,
  journal = {SIAM Journal on Computing},
  volume = {8},
  number = {3},
  pages = {410--421},
  issn = {0097-5397},
  doi = {10.1137/0208032},
}

@article{virtanenSciPy10Fundamental2020,
  title = {{{SciPy}} 1.0: Fundamental Algorithms for Scientific Computing in {{Python}}},
  shorttitle = {{{SciPy}} 1.0},
  author = {Virtanen, Pauli and Gommers, Ralf and Oliphant, Travis E. and Haberland, Matt and Reddy, Tyler and Cournapeau, David and Burovski, Evgeni and Peterson, Pearu and Weckesser, Warren and Bright, Jonathan and {van der Walt}, St{\'e}fan J. and Brett, Matthew and Wilson, Joshua and Millman, K. Jarrod and Mayorov, Nikolay and Nelson, Andrew R. J. and Jones, Eric and Kern, Robert and Larson, Eric and Carey, C. J. and Polat, {\.I}lhan and Feng, Yu and Moore, Eric W. and VanderPlas, Jake and Laxalde, Denis and Perktold, Josef and Cimrman, Robert and Henriksen, Ian and Quintero, E. A. and Harris, Charles R. and Archibald, Anne M. and Ribeiro, Ant{\^o}nio H. and Pedregosa, Fabian and {van Mulbregt}, Paul},
  year = {2020},
  month = mar,
  journal = {Nature Methods},
  volume = {17},
  number = {3},
  pages = {261--272},
  publisher = {Nature Publishing Group},
  issn = {1548-7105},
  doi = {10.1038/s41592-019-0686-2},
}

@article{wieSimplerQuantumCounting2019,
  title = {Simpler Quantum Counting},
  author = {Wie, Chu-Ryang},
  year = {2019},
  month = sep,
  journal = {Quantum Information and Computation},
  volume = {19},
  number = {11 \& 12},
  pages = {0967--0983},
  publisher = {Rinton Press},
  issn = {1533-7146},
  doi = {10.26421/QIC19.11-12-5},
}

@article{xiePerformanceUpperBound2025,
  title = {Performance Upper Bound of a {{Grover-mixer}} Quantum Alternating Operator Ansatz},
  author = {Xie, Ningyi and Xu, Jiahua and Chen, Tiejin and Lee, Xinwei and Saito, Yoshiyuki and Asai, Nobuyoshi and Cai, Dongsheng},
  year = {2025},
  month = jan,
  journal = {Physical Review A},
  volume = {111},
  number = {1},
  pages = {012401},
  publisher = {American Physical Society},
  doi = {10.1103/PhysRevA.111.012401},
}

@misc{zdeborovaStatisticalPhysicsHard2008,
  title = {Statistical {{Physics}} of {{Hard Optimization Problems}}},
  author = {Zdeborov{\'a}, Lenka},
  year = {2008},
  month = jun,
  number = {arXiv:0806.4112},
  eprint = {0806.4112},
  primaryclass = {cond-mat},
  publisher = {arXiv},
  doi = {10.48550/arXiv.0806.4112},
  archiveprefix = {arXiv},
}

@article{zhangGroverQAOA3SATQuadratic2024,
  title = {Grover-{{QAOA}} for 3-{{SAT}}: Quadratic Speedup, Fair-Sampling, and Parameter Clustering},
  shorttitle = {Grover-{{QAOA}} for 3-{{SAT}}},
  author = {Zhang, Zewen and Paredes, Roger and Sundar, Bhuvanesh and Quiroga, David and Kyrillidis, Anastasios and {Duenas-Osorio}, Leonardo and Pagano, Guido and Hazzard, Kaden R A},
  year = {2024},
  month = nov,
  journal = {Quantum Science and Technology},
  volume = {10},
  number = {1},
  pages = {015022},
  publisher = {IOP Publishing},
  issn = {2058-9565},
  doi = {10.1088/2058-9565/ad895c},
}

\end{document}